\documentclass[prl,twocolumn,showkeys,showpacs,superscriptaddress]{revtex4}
\usepackage{epsfig,latexsym,amssymb}

\begin{document}

\title{Finite-Size-Scaling at the Jamming Transition:  Corrections to Scaling and the Correlation Length Critical Exponent}

\author{Daniel V{\aa}gberg}
\affiliation{Department of Physics, Ume{\aa} University, 901 87 Ume{\aa}, Sweden}
\author{Daniel Valdez-Balderas}
\altaffiliation{Current address: School of Mechanical, Aerospace and Civil Engineering, The University of Manchester, Manchester M13 9PL, United Kingdom}
\affiliation{Department of Physics and Astronomy, University of Rochester, Rochester, NY 14627}
\author{M.~A.~Moore}
\affiliation{School of Physics and Astronomy, The University of Manchester, Manchester M13 9PL, United Kingdom}
\author{Peter Olsson}
\affiliation{Department of Physics, Ume{\aa} University, 901 87 Ume{\aa}, Sweden}
\author{S. Teitel}
\affiliation{Department of Physics and Astronomy, University of
Rochester, Rochester, NY 14627}
\date{\today}

\begin{abstract}
We carry out a finite size scaling analysis of the jamming transition in frictionless bi-disperse soft core disks in two dimensions.  We consider two different jamming protocols: (i) quench from random initial positions, and (ii) quasistatic shearing.  By considering the fraction of jammed states as a function of packing fraction for systems with different numbers of particles, we determine the spatial correlation length critical exponent $\nu\approx 1$, and show that {\it corrections to scaling} are crucial for analyzing the data.  We show that earlier numerical results yielding $\nu<1$ are due to the improper neglect of these corrections.
\end{abstract}
\maketitle

Glassy behavior in condensed matter and granular systems remains a topic of considerable controversy.  In this context, the jamming of hard or soft core particles at zero temperature has been the focus of much recent effort.  As the packing fraction $\phi$ of a granular material increases, the system undergoes a sharp jamming transition from a fluid-like state to a rigid but {\it disordered} solid state \cite{Jaeger}.  It has been proposed that this $T=0$ transition is described by a critical point, with scaling behavior similar to that at a continuous phase transition as found in equilibrium systems \cite{LiuNagel}.  A key signature of a continuous transition is a correlation length $\xi$ that diverges at the jamming $\phi_J$, $\xi\sim |\phi-\phi_J|^{-\nu}$.  Determination of the critical exponent $\nu$ is thus a key goal in establishing and characterizing the critical nature of the jamming transition.  

While it has been suggested that the value of $\nu$ is independent of the dimensionality of the system, or the specific force law between particles \cite{OHern}, the  precise numerical value of $\nu$ varies widely throughout the literature.
From theoretical consideration of soft vibrational modes in the jammed solid, Wyart et al. \cite{Wyart} argued for $\nu=1/2$.  Numerical simulations of vibrational modes led Silbert et al. in two (2D) and three (3D) dimensions \cite{Silbert} to postulate diverging transverse and longitudinal correlation lengths with exponents $\nu_T\approx0.24$ and $\nu_L\approx0.48$ respectively. $k$-core percolation models, in mean field theory, also yield \cite{Schwarz} two exponents $\nu^*=1/4$ and $\nu^\#=1/2$, while a field theoretic approach \cite{Henkes}  to jamming in 2D gave $\nu=1/4$.  Simulations by Drocco et al. \cite{Drocco} of a trace particle dragged through an incipient 2D jammed liquid resulted in a value $\nu=0.71\pm0.12$, while from a numerical finite size scaling analysis of mechanically stable states in 2D and 3D O'Hern et al. \cite{OHern} found $\nu=0.71\pm 0.08$.  A scaling analysis of velocity correlations in simulated 2D shear driven flow by two of us \cite{Olsson} previously reported that $\nu=0.6\pm0.1$.  Hatano  \cite{Hatano} obtained $\nu=0.73\pm0.05$ from simulations of shear relaxation in 3D, while relaxation of random initial states to mechanical equilibrium in 2D led Head \cite{Head} to $\nu=0.57\pm 0.05$.  Heussinger and Barat \cite{Heussinger} estimate $\nu=0.8-1.0$ from displacement correlations in a 2D system under quasistatic shearing, while Heussinger et al. \cite{Heussinger2} find a dynamic correlation length in 2D with exponent $\nu=0.9$.  Establishing the precise value of $\nu$ and determining whether all these correlations lengths are the same thus remains a crucial theoretical objective.

In this work we present a detailed finite-size-scaling analysis of the jamming transition in frictionless bi-disperse soft core disks in 2D.  Only through such a scaling analysis can one hope to clearly establish the singular behavior of the system in the limit of infinite size, and the value of critical exponents.  An advantage of the finite-size-scaling method is that it allows one to compute the  exponent $\nu$ of the most divergent length scale without the need to explicitly calculate the correlation length $\xi$ itself.

We consider two different jamming ensembles: (i) quench from random initial positions (RAND), and (ii) quasistatic shearing (QS) \cite{Heussinger}.  By considering the fraction of jammed states $f$ as a function of packing fraction $\phi$ for systems with different numbers of particles $N$, we demonstrate that the correlation length critical exponent in both ensembles is  $\nu\approx 1$.  We further show that {\it corrections to scaling} are crucial for understanding our data, and argue that earlier numerical results yielding $\nu\approx 0.7$ are due to the improper neglect of these corrections.  Our results suggest that corrections to scaling may be important in other scaling analyses of critical behavior at jamming, for example in rheological behavior.

Our model is a 50:50 bi-disperse mixture of disks with diameters in the ratio 1.4 \cite{OHern}.  Particles interact with a soft core harmonic repulsion,
\begin{equation}
V(r_{ij})=\left\{ 
\begin{array}{ll}\epsilon(1-r_{ij}/d_{ij})^2/2 &{\rm for}\quad r_{ij}<d_{ij}\\ 0 &{\rm for}\quad r_{ij}\ge d_{ij}\end{array}\right.
\end{equation}
where $r_{ij}$ is the distance between the centers of two particles $i$ and $j$, and $d_{ij}$ is the sum of their radii. Length is in units such that the smaller diameter is unity, and energy is in units such that $\epsilon=1$.  A system of $N$ disks in an area $A$ thus has a packing fraction (density)
\begin{equation}
\phi = N\pi(0.5^2+0.7^2)/(2A)\enspace.  
\label{erho}
\end{equation}

We define our two ensembles as follow.  (i) RAND: This is the ensemble introduced by O'Hern et al. \cite{OHern}.  We start with a fixed number of particles, $N$, at density $\phi$, in a square box with periodic boundary conditions.  Particles are put at random initial positions, and then a conjugate gradient method is used to relax the system to the nearest local energy minimum.  The minima resulting from many such initial configurations (we use typically 10000 for each value of $\phi$) defines the ensemble. (ii) QS: At a fixed $N$ and $\phi$, we start the system in a random initial configuration, and then apply a small shear strain step $\Delta\gamma$ using Lees-Edwards boundary conditions \cite{LeesEdwards}.  A conjugate gradient method then relaxes  the system to the nearest local energy minimum, before the system is strained again by $\Delta\gamma$.  The set of states obtained after the energy minimization, after a long total strain $\gamma$, defines the ensemble.  We choose the strain step  small enough that our results do not depend on the value of $\Delta\gamma$. For our biggest systems we use $\Delta\gamma=10^{-5}$.  We average over $10 - 20$ independent runs, each sheared a total strain $\gamma\sim 4 - 8$; for our smaller sizes, we use $\gamma$ up to $200$.

In both ensembles, we stop the energy minimization when one of the following conditions is met (i) the relative decrease in the energy after 50 iterations is smaller than $10^{-10}$, or (ii) the average energy per particle is $E/N < 10^{-16}$.  In the latter case, we consider the resulting configuration to be {\it unjammed}.  The key quantity in our analysis will be the fraction of {\it jammed} states  in the ensemble at a given value of density, $f(\phi)$.  We have verified that the energy bound (ii) gives a clear separation between the jammed and unjammed states up to the largest system size we have studied.  Further details of our numerical procedures may be found in Ref.~\cite{VOT}.

\begin{figure}
\begin{center}
\includegraphics[width=3.5in]{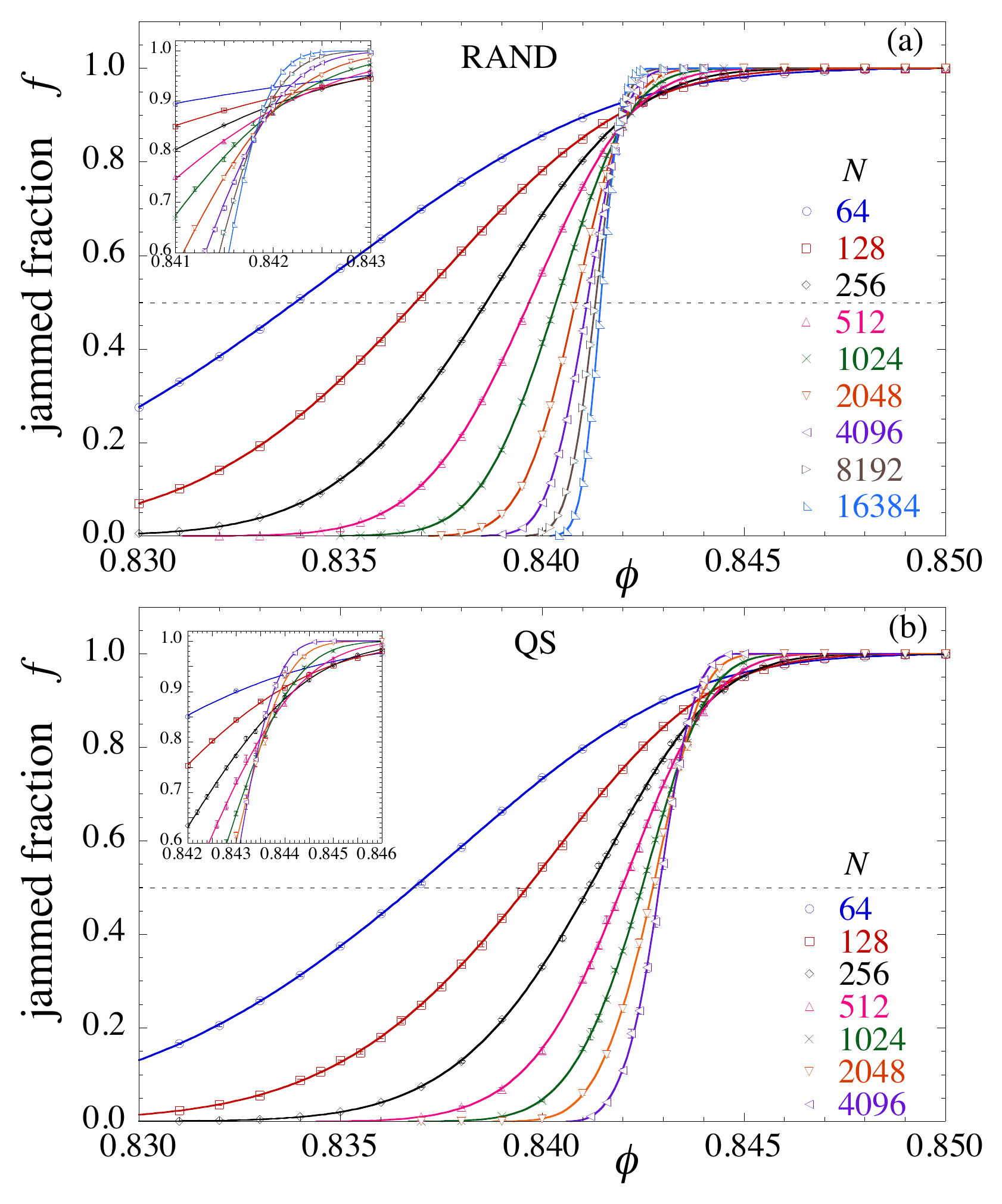}
\caption{(color online) Fraction of jammed states $f$ vs packing fraction $\phi$, for systems with number of particles $N$. (a) is the RAND ensemble, (b) is the QS ensemble.  Insets show a blow up of the region where curves for different $N$ intersect.}
\label{f1}
\end{center}
\end{figure}

In Fig.~\ref{f1} we present our results for $f(\phi)$ for systems of varying number of particles $N$ for both RAND and QS.  We see that $f(\phi)$ sharpens up and approaches a step function in the limit $N\to\infty$; this singularity in $f(\phi)$ as $N\to\infty$ is characteristic of a quantity that has {\it scaling dimension zero}.   We would thus expect, to  {\it leading} order, the finite-size-scaling behavior,
\begin{equation}
f(\phi, L) = {\cal F}\left(\delta\phi L^{1/\nu}\right)\quad{\rm where}\quad \delta\phi\equiv \phi-\phi_J\enspace,
\label{e1}
\end{equation}
$\phi_J$ is the jamming density in the thermodynamic limit $N\to\infty$, $\nu$ is the correlation length critical exponent, and $L\equiv\sqrt{N}$ is a measure of the linear size of the system. A key prediction of Eq.~(\ref{e1}) is that at $\phi=\phi_J$, curves of $f(\phi,L)$ for different $L$ should all intersect, having the common value ${\cal F}(0)$; plotting $f(\phi, L)$ vs $\delta\phi L^{1/\nu}$, curves of different $L$ should collapse to a common scaling curve.
However careful inspection of our results in Fig.~\ref{f1} (see insets) show that there is no common intersection point for the $f(\phi,L)$.  This observation leads us to conclude that, for the sizes studied here, {\it corrections to scaling} must be included in our analysis.  

We can include such corrections to scaling by generalizing Eq.~(\ref{e1}) to,
\begin{equation}
f(\phi, L) = {\cal F}_0\left(\delta\phi L^{1/\nu}\right)+L^{-\omega}{\cal F}_1\left(\delta\phi L^{1/\nu}\right)\enspace.
\label{e2}
\end{equation}
In the renormalization group framework for equilibrium critical phenomena, such corrections to scaling arise from a Taylor series expansion of the free energy in the leading irrelevant scaling field, whose scaling dimension is $-\omega$ \cite{Hasenbusch}.  We will define $f_c\equiv {\cal F}_0(0)$ as the critical value of the jamming fraction at $\phi_J$ in the limit $L\to\infty$.

One of the consequences of Eq.~(\ref{e2}) is that the functions $f(\phi,L)$ approach the $L\to\infty$ limiting step function at different rates, depending on the value of $f$.   If we define $\phi_{\bar f}(L)$ as the value of $\phi$ where $f(\phi,L)={\bar f}$, then sufficiently close to $\phi_J$ we can expand the scaling functions in Eq.~(\ref{e2}) to linear order in $\delta\phi$ to obtain,
\begin{equation}
\phi_f(L) =\phi_J- L^{-1/\nu}\left[c_{0}\delta f - (c_1-c_2\delta f) L^{-\omega}\right]\enspace,
\label{e3}
\end{equation}
where $c_0,c_1,c_2$ are constants and $\delta f \equiv f-f_c$.

\begin{figure}
\begin{center}
\includegraphics[width=3.5in]{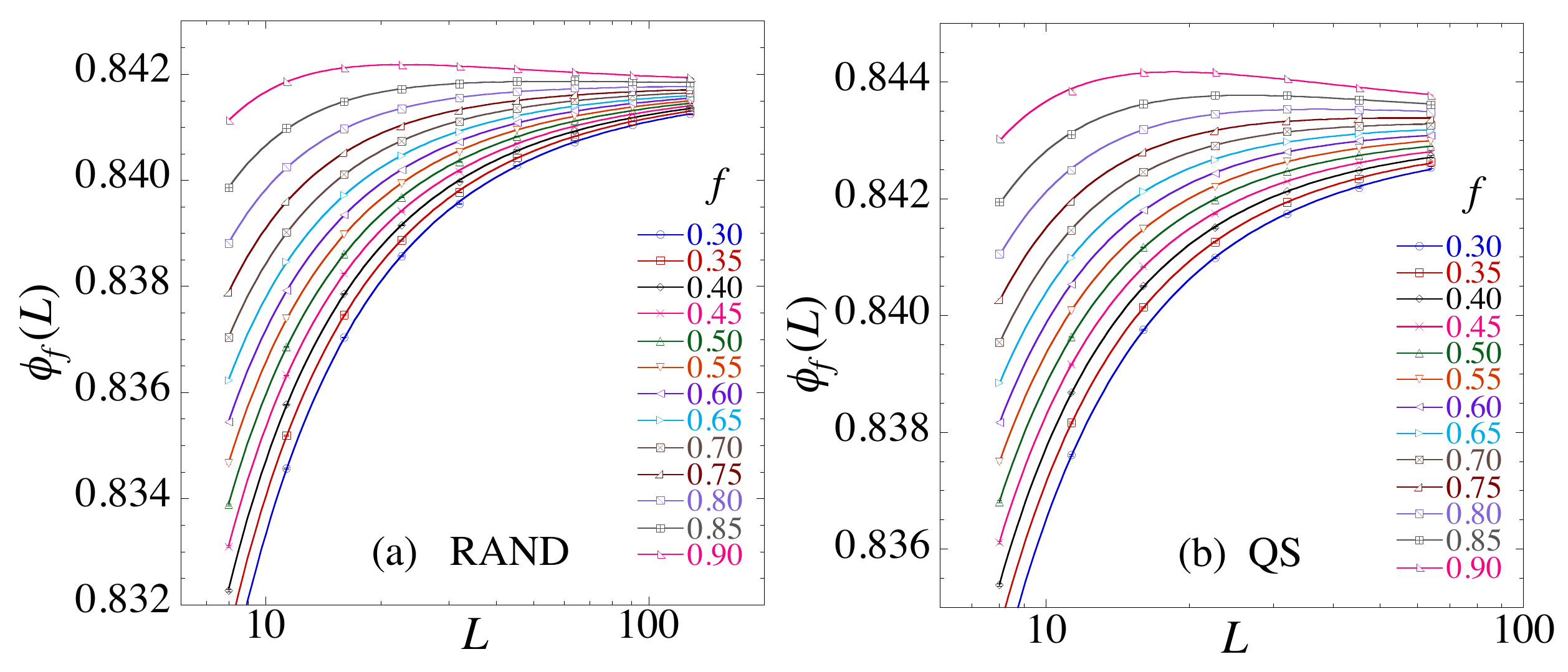}
\caption{(color online)  $\phi_f(L)$ vs $L$ for different values of $f$ for (a) RAND and (b) QS. Values of $f$ increase from bottom to top.
}
\label{f2}
\end{center}
\end{figure}

In Fig.~\ref{f2} we plot $\phi_f(L)$ vs $L$ for several values of $f$.
To interpolate between our data points so as to define the values $\phi_f(L)$, we use the following procedure.  We transform to a new variable $F\equiv \ln [f/(1-f)]$ and fit $F(\phi)$ to a fifth order polynomial over the range $|F|\le 5$.  The result gives the solid lines in Fig.~\ref{f1}.  
We see that as $f$ increases, $\phi_f(L)$ becomes non-monotonic, a clear signature of the change in sign of the leading term $L^{-1/\nu}$ in Eq.~(\ref{e3}) as $f$ increases above $f_c$.  We see that $\phi_J\approx  0.8415$  for RAND, while $\phi_J\approx 0.843$ for QS.  

\begin{figure}
\begin{center}
\includegraphics[width=3.5in]{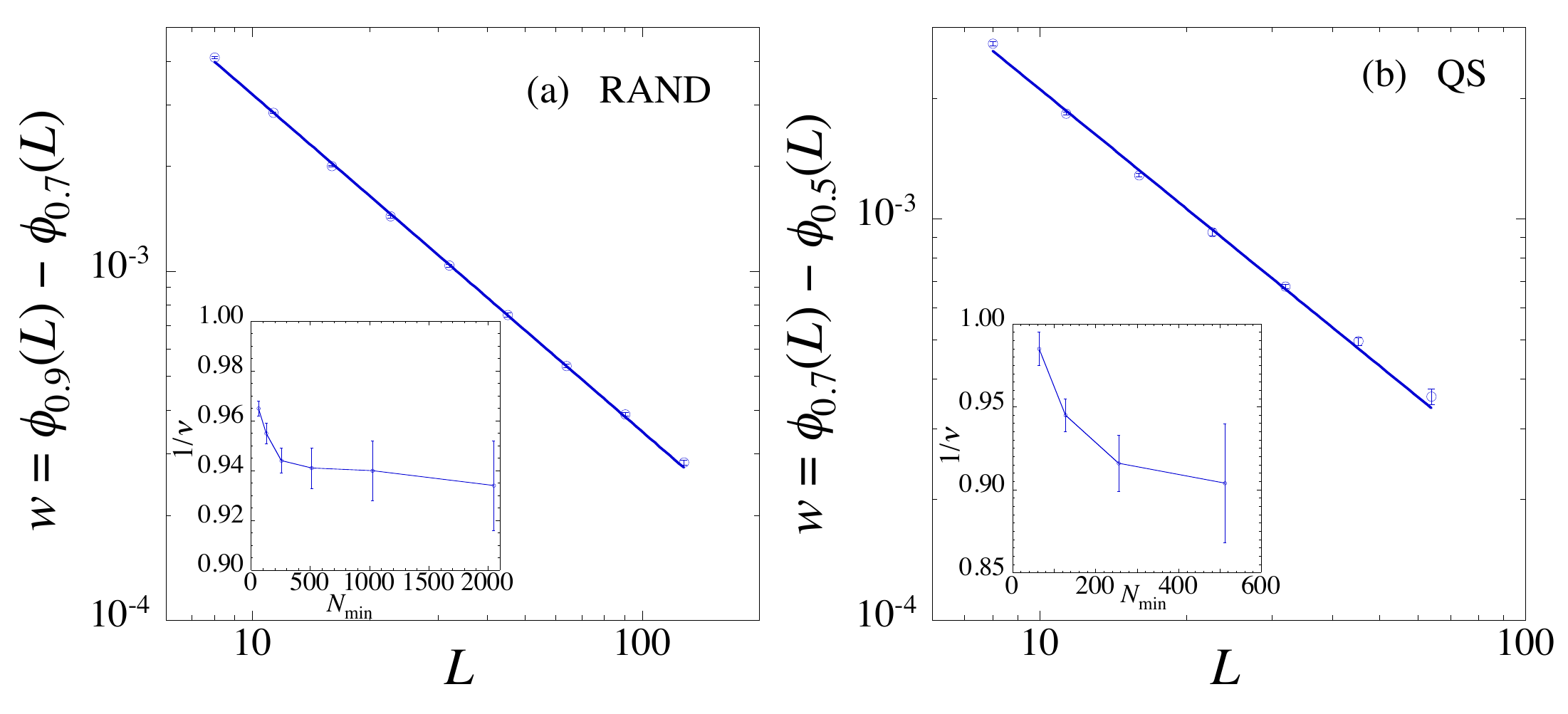}
\caption{(color online) Width $w(L)\equiv \phi_{f_2}(L)-\phi_{f_1}(L)$ vs $L$ for (a) RAND and (b) QS.  We choose $f_1$ and $f_2$ symmetrically about $f_c$; for RAND $f_1=0.7$, $f_2=0.9$; for QS $f_1=0.5$, $f_2=0.7$.  Straight line is a fit to $w\sim L^{-1/\nu}$ including all data.  Insets show the fitted value of $1/\nu$ as the minimum size system included in the fit, $N_{\rm min}$, is varied.
}
\label{f3}
\end{center}
\end{figure}

We consider next a determination of the exponent $\nu$ via Eq.~(\ref{e3}).  To eliminate the imprecisely known value of $\phi_J$, and to reduce the contribution from the correction to scaling given by $c_1$, we consider the difference,
\begin{equation}
w(L)\equiv \phi_{f_2}(L)-\phi_{f_1}(L)=aL^{-1/\nu}\left(1+bL^{-\omega}\right)\enspace,
\label{e4}
\end{equation}
where both $a$ and $b$ are proportional to $f_2-f_1$.  We choose $f_1$ and $f_2$ symmetrically about $f_c$ (with $f_c$ as determined below), and plot $w(L)$ vs $L$ for RAND and QS in Fig.~\ref{f3}.    We expect the correction term $\sim L^{-\omega}$ to get smaller, and become negligible, as $L$ increases.
We therefore ignore the correction term and fit the data to $w\sim L^{-1/\nu}$ to get the solid line in Fig.~\ref{f3}.  The insets show the resulting value of $1/\nu$ as we drop successively smaller system sizes from the fit, fitting systems of size $N_{\rm min}$ to $N_{\rm max}$ ($N_{\rm max}=16384$ for RAND, $N_{\rm max}=4096$ for QS, $L=\sqrt N$).  As expected, the value of $1/\nu$ saturates to a constant as $N_{\rm min}$ increases and the effects of the correction term become negligible.  We find from these fits the values $1/\nu= 0.93\pm 0.02$ for RAND, and $1/\nu= 0.91\pm 0.02$ for QS.  If we then fit the data for all sizes to the full Eq.~(\ref{e4}), including the correction term, we get values of $1/\nu$ consistent with those above, however the estimated error in $\omega$ is too large to determine $\omega$ to any accuracy.

\begin{figure}
\begin{center}
\includegraphics[width=3.5in]{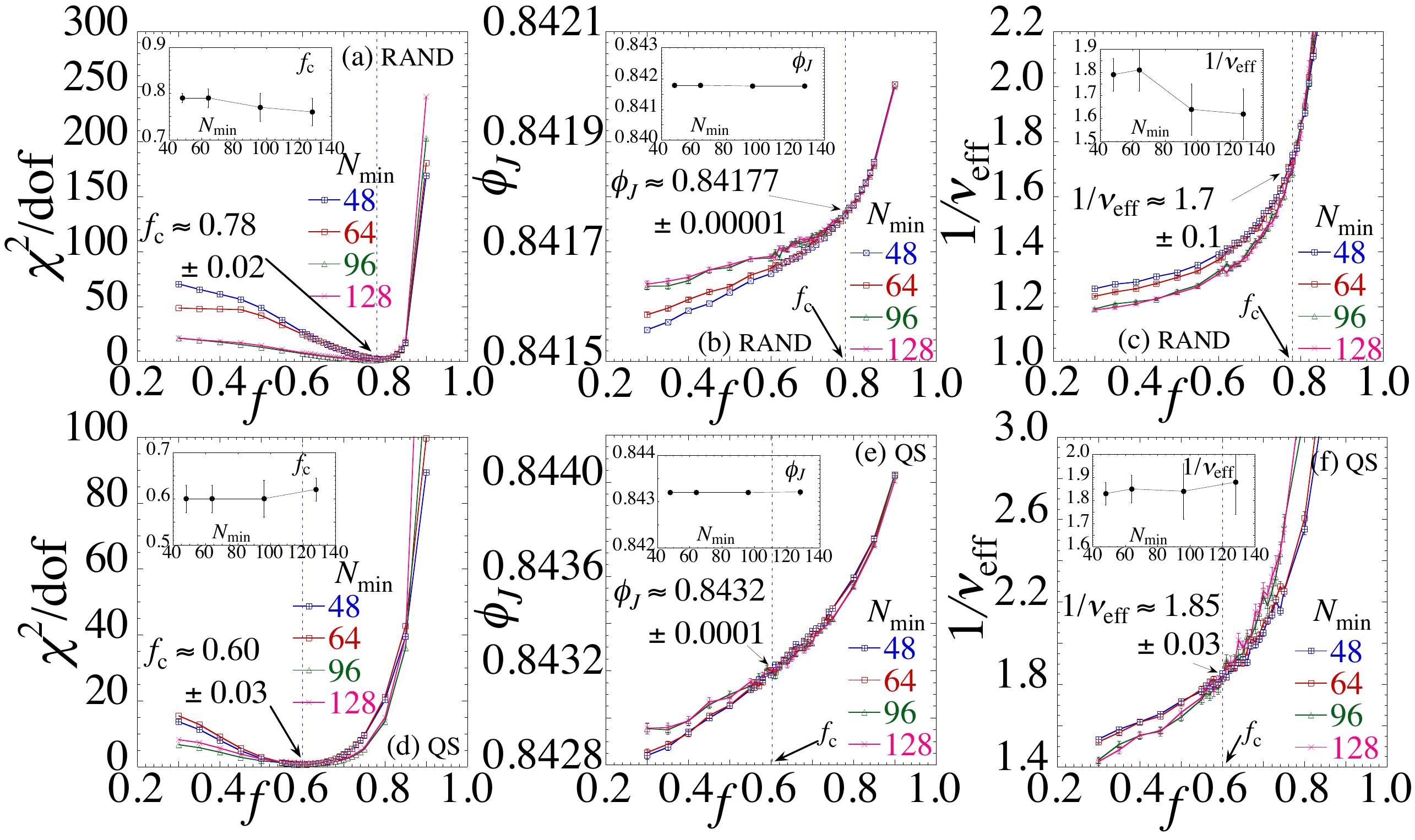}
\caption{(color online) Results from fitting data of Fig.~\ref{f2} to $\phi_f(L)=\phi_J-c L^{-1/\nu_{\rm eff}}$, using system sizes $N_{\rm min}$ to $N_{\rm max}$. We show results for $N_{\rm min}=48, 64, 96, 128$. Panels (a)-(c) are for RAND, panels (d)-(f) are for QS. (a), (d) is the $\chi^2/$dof of the fit (dof $=$ number of data points minus number of fitting parameters); the minimum of $\chi^2/$dof locates $f_c$, which then determines $\phi_J$, as shown in (b), (e), and the value of $1/\nu_{\rm eff}=1/\nu +\omega$, as shown in (c), (f).  Insets show the dependence of the fitted parameters on the value of $N_{\rm min}$.
}
\label{f4}
\end{center}
\end{figure}

To determine $\omega$, and get a more accurate value for $\phi_J$, we use the following procedure.  We fit the results for $\phi_f(L)$ of Fig.~\ref{f2} to a {\it single} power law $\phi_f(L)= \phi_J - c L^{-1/\nu_{\rm eff}}$.  Since $\phi_f(L)$ has such a single power law behavior only at $f_c$, we expect that the $\chi^2$ of the fit will be smallest when $f=f_c$.  The fitted parameters at this $f_c$ then determine $\phi_J$ and the exponent $1/\nu_{\rm eff}=1/\nu+\omega$. We show the results of such fits in Fig.~\ref{f4}, where we fit to system sizes $N_{\rm min}$ to $N_{\rm max}$, for the four different cases $N_{\rm min}=48, 64, 96, 128$.  We see that as $N_{\rm min}$ increases, the low-$f$ side of the minimum in $\chi^2$ gets increasingly shallow.  This is not surprising since the size of the correction term, and hence its effect on the fits, gets progressively smaller as $N$ increases.
Nevertheless we find quite stable values of the fitted parameters as $N_{\rm min}$ varies.
we find  $f_c=0.78\pm0.02$, $\phi_J=0.84177\pm 0.00001$, $1/\nu+\omega=1.7\pm 0.1$ for RAND, and $f_c=0.60\pm 0.03$, $\phi_J=0.8432\pm 0.0001$, $1/\nu+\omega=1.85\pm0.03$ for QS.  Combining with our earlier results for $1/\nu$ we get $\omega=0.8\pm 0.1$ for RAND and $\omega=0.94\pm0.05$ for QS.  The solid lines in Fig.~\ref{f2} result from fits to Eq.~(\ref{e3}) where we have fixed $1/\nu$ and $1/\nu+\omega$ to the values found from the analyses of Figs.~\ref{f3} and \ref{f4}.  

It is interesting to compare our results against the finite size scaling analysis of O'Hern et al. \cite{OHern}, who considered the RAND ensemble.  In that work, the authors considered the distribution $P(\phi,L)=df(\phi,L)/d\phi$, the probability density for a system of size $L$ to have its particular jamming density at $\phi$.  By considering how the location $\phi_0(L)$ of the peak in $P(\phi,L)$ approached its $L\to\infty$ limit $\phi_J$, the authors defined the critical exponent $``\nu"$ by, $\phi_J-\phi_0\sim L^{-1/``\nu"}$, and found the value $``\nu"=0.71\pm0.08$.  In terms of our analysis, we see that $\phi_0$ is the same as our $\phi_{f_0}$, where $f_0$ locates the steepest slope of $f(\phi,L)$, and $``\nu"$ is just our $\nu_{\rm eff}$.  In light of corrections to scaling, we see that $``\nu"$ should {\it not} be identified as the correlation length exponent; it is an effective exponent that arises from fitting $\phi_0$ to single power law, when the true behavior as in Eq.~(\ref{e3}) is governed by two different power laws with exponents $1/\nu$ and $1/\nu+\omega$.  If we take $f_0=0.5$, our Fig.~\ref{f4}c for the case $N_{\rm min}=64$ gives $1/\nu_{\rm eff}=1.32$, or $\nu_{\rm eff}=0.76$, in good agreement with the value found by O'Hern et al.

O'Hern et al. similarly define the full width at half maximum of $P(\phi,L)$, $w(L)$, and find the scaling $w\sim N^{-\Omega}\sim L^{-2\Omega}$, with $\Omega=0.55\pm 0.03$ or $2\Omega=1.10\pm 0.06$.   With suitable choices of $f_1$ and $f_2$, this $w$ is the same as our $w$ of Eq.~(\ref{e4}), and hence we expect for asymptotically large $N$ (where corrections to scaling become negligible) to find $2\Omega=1/\nu$.  If we assume a Gaussian form for $P(\phi,L)$ then the full width criterion corresponds to $f_1=0.124$ and $f_2=0.876$.  Computing this $w$ and fitting using sizes $N=64$ to $4096$, the same range as O'Hern et al., we get $2\Omega= 1.035\pm 0.002$, in agreement with O'Hern et al. within their estimated errors.  However if we use up to our largest size $N=16384$, then increase $N_{\rm min}$, we find $\Omega$ to systematically decrease, becoming $2\Omega=0.97\pm 0.01$ when $N_{\rm min}=2048$.  Our result remains larger than the $1/\nu=0.93$ found in Fig.~\ref{f3}, perhaps because $f_1$ is so far from $f_c$ that additional corrections to scaling arise from non-linearities in the scaling functions.  Thus we conclude that it is O'Hern et al.'s $1/(2\Omega)$ that is asymptotically the correlation length exponent $\nu$, rather than their $``\nu"$ (our $\nu_{\rm eff}$), and that their value for $2\Omega$ is larger than our $1/\nu$ due to their more limited range of sizes and their neglect of corrections to scaling. 

In other recent work \cite{OlssonTeitel}, it is found that corrections to scaling must similarly be included to properly describe the critical scaling of rheology under applied shear strain rate $\dot\gamma$, for rates of the size typically used in simulations.  The value $\phi_J=0.8415$ for shear driven jamming that was reported in earlier work by two of us \cite{Olsson}, is lower than the corresponding $\phi_J=0.8432$ found for QS here, due the neglect in that work of corrections to scaling.  Similarly, the low value $\nu=0.6$ reported in that work also results from the earlier failure to include corrections to scaling.  We expect that other numerically reported values of $\phi_J$ and $\nu$ may similarly be inaccurate due to the neglect of corrections to scaling in the analysis.

To conclude, we have demonstrated that, for the sizes $N$ typically used in simulations, including corrections to scaling is crucial for a proper description of the critical behavior at jamming, in both RAND and QS ensembles.  
Although we know no apriori reason why this should be so, it is interesting to note that corrections to scaling are similarly important in spin glass problems, another system in which the ``ordered" state appears spatially random \cite{Hasenbusch}.
Within our estimated accuracy we find the correlation length exponent $\nu$ and the correction to scaling exponent $\omega$ to be  roughly equal for the two ensembles.  We find $\omega= 0.89\pm0.12$, and $1/\nu= 0.92\pm 0.02$, or $\nu= 1.09\pm 0.02$. While the estimated statistical error in $\nu$ is small \cite{error}, our range of system sizes $L$ is not sufficiently large for us rule out the possibility that systematic errors, for example from additional or higher order corrections to scaling, could slightly alter the value of these exponents to $\nu=\omega=1$.

This work was supported by Department of Energy Grant No. DE-FG02-06ER46298, Swedish Research Council Grant No. 2007-5234, a grant from the Swedish National Infrastructure for Computing (SNIC) for computations at HPC2N and the University of Rochester Center for Research Computing.

\end{document}